\documentclass[12pt]{article}
\usepackage[utf8]{inputenc}
\usepackage{amsmath, amssymb}
\usepackage{graphicx}
\usepackage{float}
\usepackage{cite}
\usepackage{hyperref}
\usepackage{authblk}
\usepackage{geometry}
\usepackage{color}
\title{\textbf{Advancing Machine Learning Optimization of Chiral Photonic Metasurface: \\
Comparative Study of Neural Network and Genetic Algorithm Approaches}}

\author[1]{Davide Filippozzi}
\author[2]{Alexandre Mayer}
\author[2]{Nicolas Roy}
\author[3]{Wei Fang}
\author[1]{Arash Rahimi-Iman}
\affil[1]{I. Physikalisches Institut and Center for Materials Research, Justus-Liebig-Universität Gießen, D-35392 Giessen, Germany}
\affil[2]{Department of Physics, Namur Institute for Complex Systems (naXys), University of Namur, B-5000 Namur, Belgium}
\affil[3]{College of Optical Science and Engineering, Zhejiang University, Hangzhou 310027, China}
\date{\today}

\begin{document}

\maketitle

\begin{abstract}
Chiral photonic metasurfaces provide unique capabilities for tailoring light-matter interactions, which are essential for next-generation photonic devices. 
Here, we report an advanced optimization framework that combines deep learning and evolutionary algorithms to significantly improve both the design and performance of chiral photonic nanostructures. 
Building on previous work utilizing a three-layer perceptron reinforced learning and stochastic evolutionary algorithm with decaying changes and mass extinction for chiral photonic optimization, our study introduces a refined pipeline featuring a two-output neural network architecture to reduce the trade-off between high chiral dichroism (CD) and reflectivity. Additionally, we use an improved fitness function, and efficient data augmentation techniques. 
A comparative analysis between a neural network (NN)-based approach and a genetic algorithm (GA) is presented for structures of different interface pattern depth, material combinations, and geometric complexity. We demonstrate a twice higher CD and the impact of both the corner number and the refractive index contrast at the example of a GaP/air and PMMA/air metasurface as a result of superior optimization performance. Additionally, a substantial increase in the number of structures explored within limited computational resources is highlighted, with tailored spectral reflectivity suggested by our electromagnetic simulations, paving the way for chiral mirrors applicable to polarization-selective light–matter interaction studies.
\end{abstract}

\section{Introduction}

Nanophotonic structures with tailored optical properties have emerged as a cornerstone for advanced applications in imaging, optoelectronics, quantum information, and biome\-dicine~\cite{alias2018review,hughes2018adjoint,javaid2024reviewing}. 
A particular area of interest is the design of \textit{chiral photonic metasurfaces}, which exhibit distinct responses to left- and right-circularly polarized light (LCP and RCP). Such structures play a crucial role in valleytronics~\cite{guddala2019valley}, chiral molecule detection~\cite{im2024perspectives}, and control of optical angular momentum~\cite{kuhner2023unlocking}.

The manual design of chiral metasurfaces remains highly challenging due to unintuitive geometric relationships and complex electromagnetic interactions. 
Algorithmic optimization techniques, such as inverse design~\cite{li2022empowering,Wiecha_2021} based on auto-differentiation~\cite{hughes2019forward,minkov2020inverse} or neural-adjoint methods~\cite{deng2021neural}, have been developed to address these challenges. Machine learning (ML) and deep learning (DL) have shown great potential to accelerate photonic structure optimization by replacing time-consuming iterative simulations with predictive models~\cite{ma2019probabilistic,Jiang2019,So_2021,So_2023,ma2021deep}. These DL techniques are used as surrogates for the computational-expensive optical simulations, but also to approximate the inverse relation between the desired output and the geometric/material input parameters\cite{Fang2024inverse}. Once trained on a dataset made of a limited number of optical simulations (data augmentation used when possible), these algorithms can explore the design space at a much lower cost. These algorithms often involve autoencoders to work with reduced lattent-space representations, UNets for 2-D map transformations\cite{Roy2024pra} and/or adversarial techniques to improve the generation of candidate solutions\cite{Christensen_2020,Wiecha_2021}. A recent trend is to incorporate physics-informed modules in these algorithms to constrain the exploration to solutions that respect relevant physical equations\cite{Chen2020PINN,Lim2022MaxwellNet,Deng2025}. Large Language Models are finally also used for the design of metamaterials with on-demand properties\cite{LuJordan2025agentic}.

Our previous study~\cite{mey2022machine} demonstrated that both neural networks and evolutionary algorithms can be applied to optimize chiral metasurfaces, leading to designs with increased circular dichroism. 
However, this initial proof-of-principle left several open questions, namely how neural network accuracy and generalization can be improved for complex geometries, what computational trade-offs arise when comparing neural-network and genetic-algorithm approaches, and whether hybrid strategies can effectively combine the strengths of both methods.

Here, we present a follow-up study addressing these key questions by introducing an advanced ML optimization pipeline. 
Our pipeline is designed to achieve low computational cost, high scalability, and rapid feasibility testing of unintuitive chiral structures. 
The ultimate application goal is to realize lithographically feasible chiral mirrors for selective light–matter interaction experiments involving 2D materials such as tungsten disulfide (WS$_2$).
The work concludes with a discussion of our improvements, the comparison of results obtained for GaP and PMMA based on both distinct approaches and the strengths of each method for such class of optimization problems.

\begin{figure}[H]
\centering
\includegraphics[width=1\linewidth]{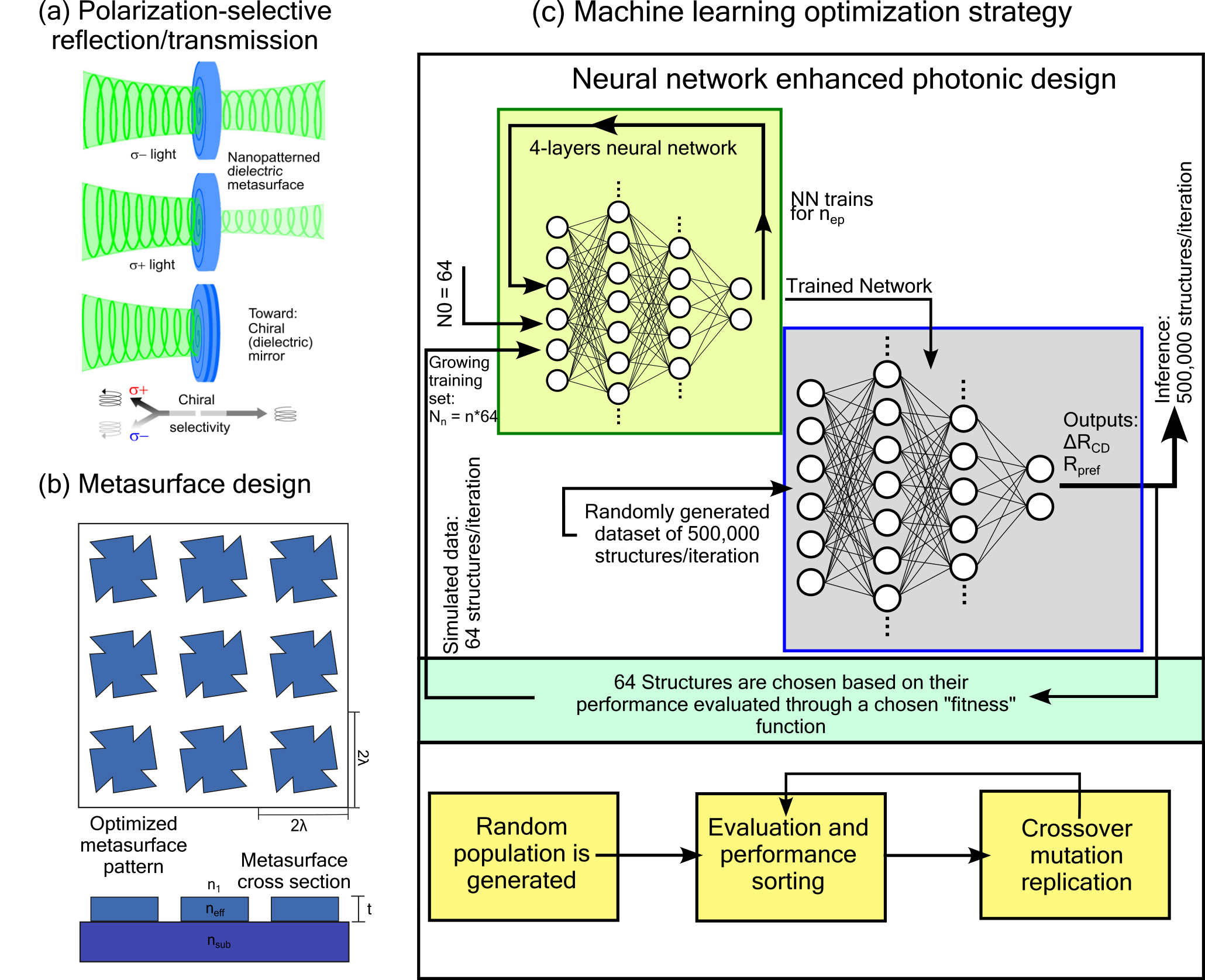}
\caption{Sketched chiral metasurface optimization task: (a) the polarization selective reflectance, (b) The geometry (top view unit cell, lattice and cross section), and (c) two distinct algorithms for the inference of a high CD and suitable spectral reflectivity performance of the patterned air--substrate interface (top: neural-network algorithm; bottom: evolutionary algorithm). Drawings after Ref. \cite{mey2022machine}.}
\label{fig:overview1}
\end{figure}

\section{Methods}

To simulate the behavior of the metasurfaces, the software Khepri \cite{Khepri} was used throughout this work. Khepri implements the Fourier modal method \cite{Moharam_1981,RCWA}, which enables efficient calculation of transmission and reflectivity spectra of layered dielectric materials with in-plane periodic structures. In this method, the two-dimensional periodic distribution of the relative permittivity is approximated by a Fourier series expansion of the geometry. As a result, the accuracy of the simulation increases with the number of Fourier terms included, at the cost of higher computational requirements. Therefore, a balance between accuracy and computational
efficiency must be achieved. In this work, $n_{\rm f} = 13$ Fourier orders were used to obtain a suitable compromise between precision and computational cost. Moreover, the choice makes simulation accuracy relatively comparable to Ref.~\cite{mey2022machine}, which instead used the Fourier modal method implemented by the software S4 \cite{S4}.

\subsection{Simulation Framework} 

Structures were modeled as periodic metasurfaces with unit cells containing $n_{\rm corners}$ = 3, 4 and 5 corners (vertices) per quadrant to investigate the role of geometric complexity (C$_4$ symmetry is assumed). 
The substrate material was chosen as GaP due to its high refractive index ($n = 3.34$ at 615 nm~\cite{aspnes1983dielectric}) and low absorption in the visible range.
The simulated structures had to respect two physical constrains: the angle between the edges connection can not be lower than 45 degrees, and there can be no intersection between edges.
The entire simulation was kept unitless to simplify the design space.
The central design frequency $f_0$ and wavelenght $\lambda_0$, for which a high CD and high reflectivity shall be achieved, were both set to one.
The lattice constant $l$ of the unit cell was defined as $l=2\lambda_0$, and the structure depth was also defined as a fraction or a multiple of $\lambda_0$.
The first target quantity to optimize was the circular dichroism integral, defined as:
\begin{equation}
 \Delta R_{\rm CD} = \int_{f_{\rm low}}^{f_{\rm high}} \big[R_{\rm LCP}(f) - R_{\rm RCP}(f)\big] df,
\end{equation}
where $R_{\rm LCP}$ and $R_{\rm RCP}$ denote the reflectivity for left- and right-circularly polarized light, respectively. The integral is evaluated in the frequency range $f_{\rm low} = 0.95$ to $f_{\rm high}= 1.05$ with a stepsize of 0.002. The second target quantity is the reflectivity of the preferred polarization, defined as:
\begin{equation}
 R_{\rm pref} = \max\left( \int_{f_{\rm low}}^{f_{\rm high}} R_{\rm RCP}(f) \space df, \int_{f_{\rm low}}^{f_{\rm high}} R_{\rm LCP}(f) \space df \right).
\end{equation}
In the optimizations presented here-after, we actually seek to maximize both the absolute value of $\Delta R_{\rm CD}$ and $R_{\rm pref}$ through an effective objective function defined as $F= |\Delta R_{\rm CD}| \times R_{\rm pref}$.

\subsection{Neural Network-Based Optimization}

Initially, the neural-network pipeline was established in a particularly compact and computationally inexpensive format. Because the results of~\cite{mey2022machine} with the replaced simulation code (Khepri~\cite{Khepri} instead of $S^4$~\cite{S4}) are reproducible, in the following study, the integrated simulations delivered the basis for efficient fast computation and comparability in optimization results regarding both the neural network and genetic algorithm pipelines.

The neural network (NN) approach used an architecture consisting of 
$2\times n_{\rm corners}$ input neurons, a first hidden layer with 16 neurons, a second hidden layer with 8 neurons and 2 output neurons to simultaneously predict $\Delta R_{\rm CD}$ and $R_{\mathrm{pref}}$. The activation function used in the hidden layers is the LeakyRELU function, with a slope of 0.01.
Training utilized a dataset that increases linearly at each iteration for a total of 15 iterations. In the first iteration, 64 training examples are used. For the next iterations, 64 additional training examples are selected and added to the dataset, reaching a total of 960 in the final iteration. The optimizer used for training was RMSprop with a learning rate of 0.001, and the loss function was the mean square error.

To improve convergence and compatibility with the genetic algorithm, we used the enhanced fitness function
\begin{equation}
 F = |\Delta R_{\rm CD}| \times  R_{\mathrm{pref}}.
\end{equation}
This fitness function is used in the neural-network optimization to select the structures that are added to the training dataset at each iteration.

Several key enhancements were introduced in the optimization routine compared to the previous study, including the use of geometric data augmentation to improve model robustness, an adaptive strategy for selecting the number of training epochs to avoid both overfitting and underfitting, and an expansion of the design space achieved by varying structural thickness, material composition, and geometric complexity.

\subsection{Genetic Algorithm Optimization}

\begin{figure}[ht]
 \centering
 \includegraphics[width=0.6\linewidth]{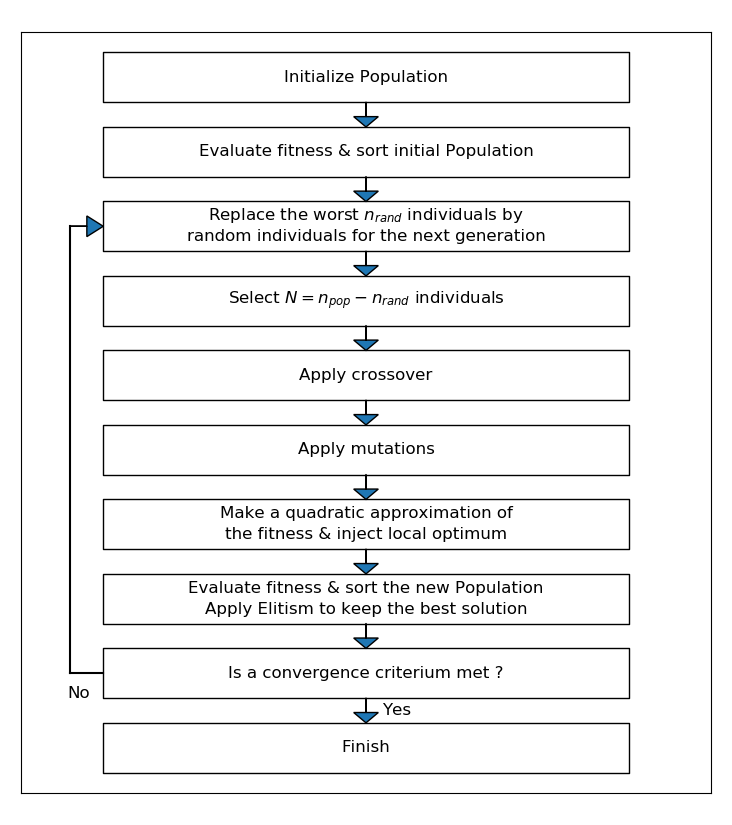}
 \caption{Workflow of the Genetic Algorithm}
 \label{GA_workflow}
\end{figure}

The Genetic Algorithm (GA) used for this study was described with details in previous work\cite{Mayer_SPIE2018,Mayer_SPIE2020,Mayer_OE2022,Mayer_SPIE2024}.
The idea consists in working with a population of $n_{\rm pop}$=50 individuals, which represent possible coordinate values for the
corners of the structures considered. The number of variables considered by the GA is hence $n=2 \times n_{\mathrm{corners}}$.
The initial population consists of random individuals that respect the imposed constraint (no angle lower than 45 degrees). The fitness $F = |\Delta R_{\rm CD}| \times R_{\mathrm{pref}}$ of each individual is then calculated with Khepri. The $n_{\rm rand}$ less fit individuals are replaced by new random individuals for the next generation ($n_{\rm rand}=0.1\times n_{\rm pop}$ initially; it decreases according to the genetic similarity of the population)\cite{Mayer_SPIE2018}.
For the rest of the population, a rank-based roulette wheel selection is used to select the best individuals (a given individual can be selected several times)\cite{Eiben_2007}. For each pair of selected individuals (the parents), we apply a crossover operator to define children for the next generation (probability of 70\%) or we leave these individuals unchanged (probability of 30\%). The children obtained by crossover are further subjected to mutations. These mutations consist of random flipping of the binary encoding used for the representation of the corner coordinates\cite{Mayer_SPIE2018}. We make sure that the individuals obtained by these crossover and mutation operators respect the imposed constraint (these operators are repeated if necessary until an acceptable solution is found). At each generation, the data collected by the Genetic Algorithm is used to establish a quadratic approximation of the fitness in the neighborhood of the best-so-far individual. The optimum of this approximation is then injected in the next generation\cite{Mayer_SPIE2018}. We finally apply elitism to make sure that the best-so-far solution is not lost when going from one generation to the next. These different steps are repeated from generation to generation until a termination criteria is met. A schematic representation of this workflow is given in Fig. \ref{GA_workflow}.

\section{Results and discussion}

With the improved pipeline, we increased the number of structures analyzed per optimization run from 1.4 million to 7.5 million, which represents approximately a fivefold improvement in computational efficiency (the number of Khepri simulations is 960 for each optimization). The NN approach achieved strong performance after 5 iterations. After this, the predictive capability of the network does not increase substantially. 

The GA established final solutions after typically 150 generations for $n_{\rm corners}=3$, 250 generations for $n_{\rm corners}=4$ and 400 generations for $n_{\rm corners}=5$ (typical numbers of generations until the final best solution was found). The GA was actually run 3 times on each case. In these repeated optimizations, the best solution established by a given run of the GA is injected in the initial population of the next run to give a chance to improve this solution.

\subsection{Neural network capability improvements and efficiency}

\begin{figure}[H]
\centering
\includegraphics[width=1\linewidth]{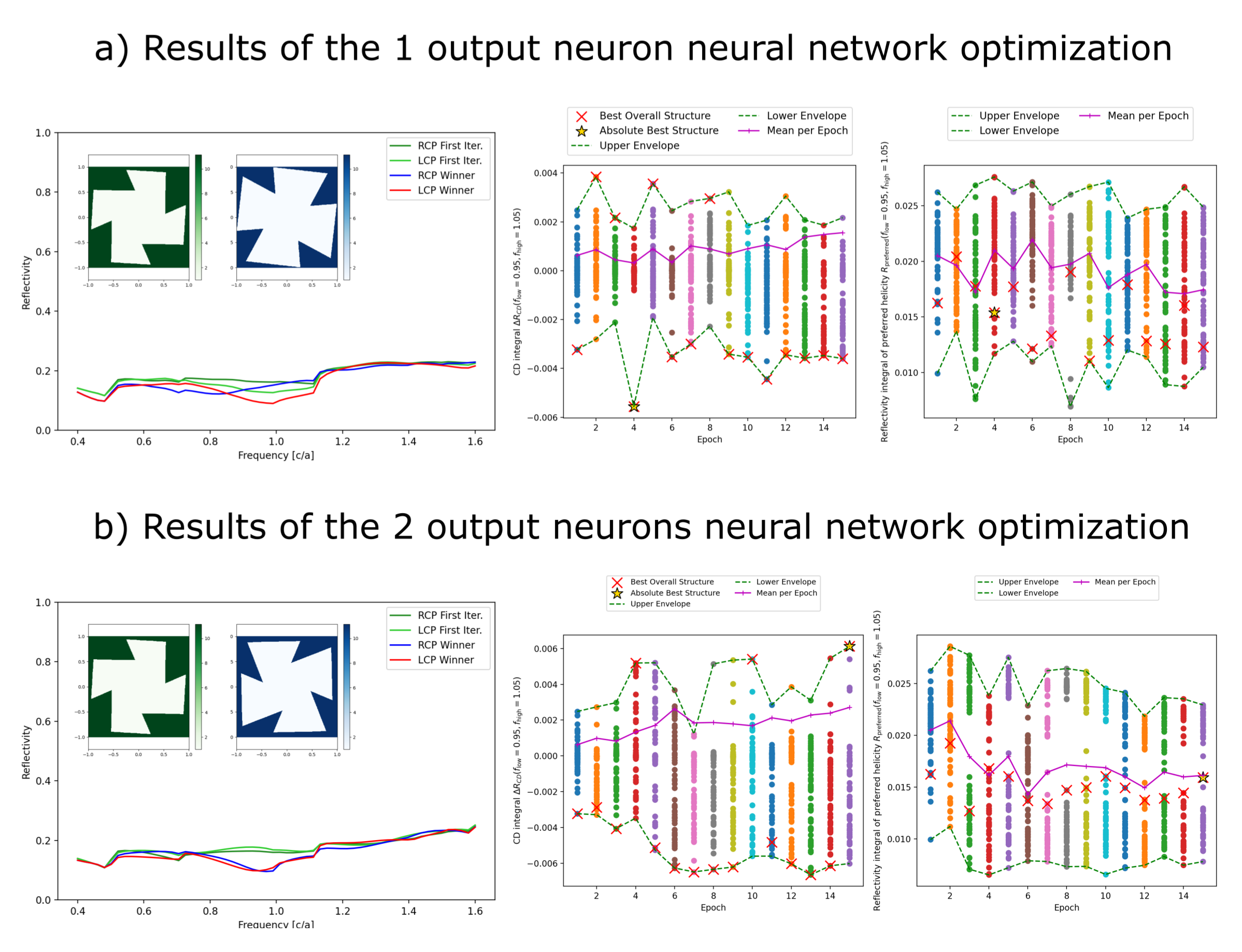}
\caption{Improvement of the few-layer perceptron reinforcement-learning pipeline: from a single output neuron (a) to two output neurons and an enhanced fitness evaluation function (b). Insets: The structures shown in green represent the best candidate identified after the first iteration (left side), during which 64 randomly generated configurations are evaluated. The structures shown in blue (right side) correspond to the optimized designs obtained at the end of the full optimization process.}
\label{fig:results1}

\subsection{Structure thickness variation and geometry complexity}\label{ThicknessAndCompl}

\end{figure}

In the case of the single-output-neuron neural network (representative of $\Delta R_{\rm CD}$), the best structure found for $n_{\rm corners}=3$
in the first iteration exhibits a $\Delta R_{\rm CD}$ of approximately -0.0032 (see Fig. \ref{fig:results1}). This result rapidly improves during the optimization, with the best structure reaching a CD of approximately -0.0056, which represents a significant enhancement. However, the network after the final iteration presents a high validation loss of around 0.5, which is considered substantial for this type of optimization problem. Despite this, the average CD value (which is the only quantity optimized at this stage) increases consistently across the 15 iterations, whereas the average reflectivity follows an opposite trend. These results are comparable to those reported previously in~\cite{mey2022machine}. The high validation loss observed in this configuration is attributed to underfitting, as the network is too simple to effectively learn the nonlinear relationship between the corner coordinates and the CD values of the structures. Furthermore, the use of the ReLU activation function in such a compact network can lead to the deactivation of some hidden neurons, further reducing model responsiveness.

To obtain structures with high CD without excessive loss in reflectivity, a second output neuron was added to the neural network. The architecture of this network is described in Section 2.2. In this configuration, the optimized structure achieved a $\Delta R_{\rm CD}$ of approximately -0.0061 and a preferred reflectivity $R_{\rm pref}$ of approximately 0.0159, thus improving both figures of merit. The validation loss was also reduced to approximately 0.1, which is five times lower than that obtained with the previous single-output architecture. These improved results arise directly from the enhancements introduced to both the network architecture and the training process.

A data-augmentation strategy exploiting the mirror symmetry between enantiomeric planar chiral metasurfaces was implemented, allowing the training dataset to be doubled without additional simulations, a visualization of the effect of this strategy on the dataset is reported in Figure \ref{fig:Dataset} of the Appendix . Reversing the handedness of a chiral structure produces a mirror-symmetric response: the reflected intensities under left- and right-circularly polarized illumination remain the same, but the associated polarization labels are interchanged because the direction of polarization rotation is reversed~\cite{bai2007optical}. This symmetry enables each simulated sample to be paired with its enantiomeric counterpart simply by swapping the LCP and RCP labels, providing an augmented dataset fully consistent with the underlying physics. This approach preserved the total computation time while producing a more evenly distributed dataset across the design space, ensuring a balanced representation of structures favoring left- and right-circularly polarized light. Additionally, an adaptive epoch selection strategy was introduced to automatically determine the optimal number of training epochs in each iteration, mitigating potential overfitting during prolonged training, an effect that becomes increasingly relevant with larger networks. Overall, the improvements applied to the NN-based optimization proved successful in achieving superior performance and efficiency.

Figure \ref{fig:results1}, and Figure \ref{fig:Validation} of the Appendix, summarise the achievements with the improved neural network architecture to minimize the trade-off between $\Delta R_{\rm CD}$ and $R_{\rm pref}$ and compares the winner structures among the two machine learning approaches.

\begin{figure}[!htbp]
    \centering
    \includegraphics[width=0.8\linewidth]{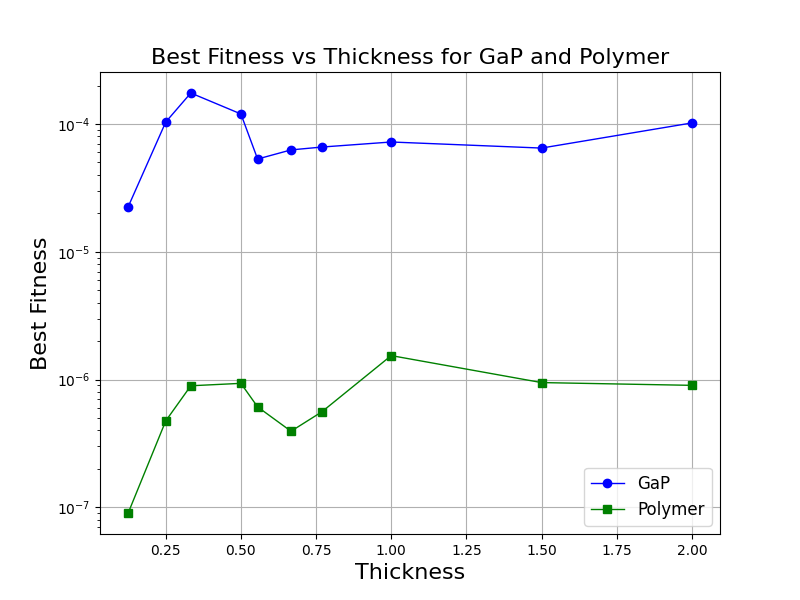}
    \caption{Results of the thickness optimization for the two tested compositions, GaP/Air in blue and PMMA/Air in green.
    Each dot represents the best individual structure found at a specific thickness. The GaP/Air composition achieved the highest fitness value at thickness $\lambda_0/3$ while the PMMA/Air composition achieved the highest fitness at thickness $\lambda_0$.}
    \label{fig:ThicknessTest}
\end{figure}

The improved neural network-based optimization framework was employed to analyze chiral structures with varying thicknesses and material compositions (GaP/Air and PMMA/Air). For each composition, ten different thicknesses were evaluated to identify the optimal value that maximizes the fitness function of the best structure found. As shown in Fig.~\ref{fig:ThicknessTest}, the optimal thickness for the GaP/Air composition was identified at $t=\lambda_0/3$, while for PMMA/Air it was found at $t=\lambda_0$. Across all thicknesses, the PMMA/Air structures exhibited fitness values nearly two orders of magnitude lower than those of their GaP/Air counterparts.

The large difference in fitness between the two materials is primarily attributed to their distinct refractive index contrasts. PMMA has a refractive index of approximately 1.5 at 615 nm, while GaP exhibits a much higher refractive index of approximately 3.34 at the same wavelength. This contrast leads to substantially reduced reflectivity from the PMMA/Air interface compared to the GaP/Air interface.

\begin{figure}[htbp]
\centering
\includegraphics[width=1.0\linewidth]{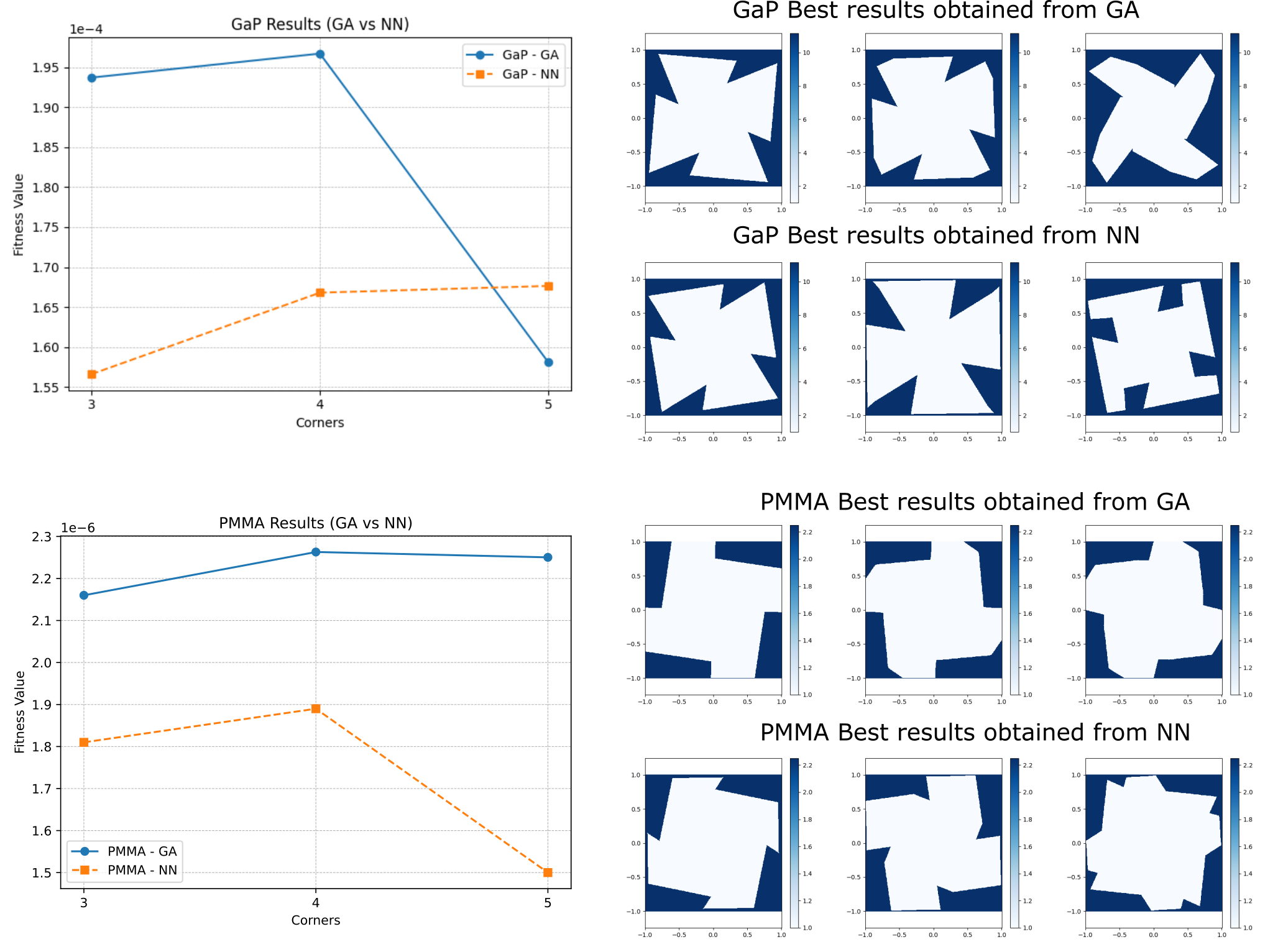}
\caption{Structures complexity increase: comparison of polarization-sensitive structure performance obtained from neural network and genetic algorithm optimizations.}
\label{fig:Structure Compl}
\end{figure}

The optimal thicknesses identified in this study were subsequently used in the analysis of structural complexity. The results are presented in Fig.~\ref{fig:Structure Compl} and Table \ref{Table_comparison}. As shown on the left panels of the figure, the GA almost consistently outperforms the neural-network-based optimization, except in the case of GaP/Air structures with five corners per quadrant, where the NN approach achieves comparable performance. On average, the GA produces structures with fitness scores approximately 15\% higher than those obtained through the NN-based optimization.

It is noteworthy that both methods produce similar shapes for the cases with three and four corners, while for five corners the resulting geometries diverge. In the NN-based optimization, the structures tend to deviate from the typical “windmill”-like shape commonly found in simpler configurations.
To achieve these results, the GA required a larger number of simulations per optimization run compared to the NN approach (see again Table \ref{Table_comparison}). In the NN pipeline, the number of simulations is fixed at 960, evenly distributed across the 15 iterations. In contrast, the GA does not have a fixed number of generations, leading to variation in the number of simulated structures. In the present study, the GA used on average 6846 simulations per run to establish final solutions for the GaP/Air composition and 4034 simulations per run for the PMMA/Air composition (the number of fitness evaluations tends to increase with the complexity of the problem). Quantitatively, the GA achieved slightly higher $|\Delta R_{\rm CD}|$ and $R_{\rm pref}$ values, whereas the NN approach excelled in computational speed and early-stage prediction accuracy.

\begin{table}[htbp]
 {\scriptsize
 \begin{center}
  %\negpar
  \begin{tabular}{|l|c|c|}
  \hline
  $n_{\rm corners}$=3  &  GA on Air/GaP ($t=\lambda_0/3$)     & NN on Air/GaP ($t=\lambda_0/3$) \cr
  \hline
  LCP   & 2.016E-02                                           & 1.901E-02 \cr
  RCP   & 1.055E-02                                           & 1.089E-02 \cr
  $2*|{\rm LCP}-{\rm RCP}|/({\rm LCP}+{\rm RCP})$ & 0.626     & 0.547 \cr
  $|{\rm LCP}-{\rm RCP}| * \max({\rm LCP, RCP})$  & 1.937E-04 & 1.566E-04 \cr
  \#iterations  & 340(250)+122(32)+241(169)                   & 15  \cr
  \#simulations & 9229(7041)+3279(1076)+5044(4141)            & 960 \cr
  \hline
%%%%%%%%%%%%%%%%%%%%%%%%%%%%%%%%%%%%%%%%%%%%%%%%%%%%%%%%%%%%%%%%%%%%%%%%%%%%%%%%%%%%%%%%%%%%%%%%
  \hline
  $n_{\rm corners}$=4  &  GA on Air/GaP ($t=\lambda_0/3$)     & NN on Air/GaP ($t=\lambda_0/3$) \cr
  \hline
  LCP   & 2.040E-02                                           & 1.785E-02\cr
  RCP   & 1.077E-02                                           & 8.501E-02\cr
  $2*|{\rm LCP}-{\rm RCP}|/({\rm LCP}+{\rm RCP})$ & 0.618     & 0.709\cr
  $|{\rm LCP}-{\rm RCP}| * \max({\rm LCP, RCP})$  & 1.967E-04 & 1.668E-04\cr
  \#iterations  & 233(229)                                    & 15  \cr
  \#simulations & 5140(5085)                                  & 960 \cr
  \hline
%%%%%%%%%%%%%%%%%%%%%%%%%%%%%%%%%%%%%%%%%%%%%%%%%%%%%%%%%%%%%%%%%%%%%%%%%%%%%%%%%%%%%%%%%%%%%%%%
  \hline
  $n_{\rm corners}$=5  &  GA on Air/GaP ($t=\lambda_0/3$)     & NN on Air/GaP ($t=\lambda_0/3$) \cr 
  \hline
  LCP   & 2.041E-02                                           & 1.961E-02 \cr
  RCP   & 1.266E-02                                           & 1.111E-02\cr
  $2*|{\rm LCP}-{\rm RCP}|/({\rm LCP}+{\rm RCP})$ & 0.469     & 0.557 \cr
  $|{\rm LCP}-{\rm RCP}| * \max({\rm LCP, RCP})$  & 1.581E-04 & 1.676E-04 \cr
  \#iterations  & 902(752)+210(183)+445(295)                  & 15  \cr
  \#simulations & 21339(18589)+4282(3952)+10973(8037)         & 960 \cr
  \hline
%%%%%%%%%%%%%%%%%%%%%%%%%%%%%%%%%%%%%%%%%%%%%%%%%%%%%%%%%%%%%%%%%%%%%%%%%%%%%%%%%%%%%%%%%%%%%%%%
  \end{tabular}
 \end{center}
%%%%%%%%%%%%%%%%%%%%%%%%%%%%%%%%%%%%%%%%%%%%%%%%%%%%%%%%%%%%%%%%%%%%%%%%%%%%%%%%%%%%%%%%%%%%%%%%
%%%%%%%%%%%%%%%%%%%%%%%%%%%%%%%%%%%%%%%%%%%%%%%%%%%%%%%%%%%%%%%%%%%%%%%%%%%%%%%%%%%%%%%%%%%%%%%%
 \begin{center}
  \begin{tabular}{|l|c|c|}
  \hline
  $n_{\rm corners}$=3  &  GA on Air/Polymer ($t=\lambda_0$)     & NN on Air/Polymer ($t=\lambda_0$) \cr
  \hline
  LCP   & 1.797E-03                                             & 1.879E-03 \cr
  RCP   & 2.622E-03                                             & 2.579E-03\cr
  $2*|{\rm LCP}-{\rm RCP}|/({\rm LCP}+{\rm RCP})$ & 0.373       & 0.314\cr
  $|{\rm LCP}-{\rm RCP}| * \max({\rm LCP, RCP})$  & 2.163E-06   & 1.805E-06 \cr
  \#iterations  & 169(79)+147(79)+171(81)                       & 15  \cr
  \#simulations & 4099(2424)+3390(2310)+4349(2512)              & 960 \cr
  \hline
%%%%%%%%%%%%%%%%%%%%%%%%%%%%%%%%%%%%%%%%%%%%%%%%%%%%%%%%%%%%%%%%%%%%%%%%%%%%%%%%%%%%%%%%%%%%%%%%
  \hline
  $n_{\rm corners}$=4  &  GA on Air/Polymer ($t=\lambda_0$)     & NN on Air/Polymer ($t=\lambda_0$) \cr
  \hline
  LCP   & 2.667E-03                                             & 2.489E-03\cr
  RCP   & 1.818E-03                                             & 1.730E-03 \cr
  $2*|{\rm LCP}-{\rm RCP}|/({\rm LCP}+{\rm RCP})$ & 0.378       & 0.360 \cr
  $|{\rm LCP}-{\rm RCP}| * \max({\rm LCP, RCP})$  & 2.263E-06   & 1.189E-06\cr
  \#iterations  & 249(129)                                      & 15  \cr
  \#simulations & 6421(3842)                                    & 960 \cr
  \hline
%%%%%%%%%%%%%%%%%%%%%%%%%%%%%%%%%%%%%%%%%%%%%%%%%%%%%%%%%%%%%%%%%%%%%%%%%%%%%%%%%%%%%%%%%%%%%%%%
  \hline
  $n_{\rm corners}$=5  &  GA on Air/Polymer ($t=\lambda_0$)     & NN on Air/Polymer ($t=\lambda_0$) \cr
  \hline
  LCP   & 2.679E-03                                             & 2.031E-03 \cr
  RCP   & 1.839E-03                                             & 2.607E-03 \cr
  $2*|{\rm LCP}-{\rm RCP}|/({\rm LCP}+{\rm RCP})$ & 0.372       & 0.248 \cr
  $|{\rm LCP}-{\rm RCP}| * \max({\rm LCP, RCP})$  & 2.250E-06   & 1.501E-06\cr
  \#iterations  & 458(308)+304(154)+296(146)                    & 15  \cr
  \#simulations & 11482(8459)+7527(4448)+7536(4244)             & 960 \cr
  \hline
%%%%%%%%%%%%%%%%%%%%%%%%%%%%%%%%%%%%%%%%%%%%%%%%%%%%%%%%%%%%%%%%%%%%%%%%%%%%%%%%%%%%%%%%%%%%%%%%
  \end{tabular}
 \end{center}
%%%%%%%%%%%%%%%%%%%%%%%%%%%%%%%%%%%%%%%%%%%%%%%%%%%%%%%%%%%%%%%%%%%%%%%%%%%%%%%%%%%%%%%%%%%%%%%%
 }
 \caption{Performance of the best solutions found by the Genetic Algorithm and the Neural Network approaches, with their respective number of iterations and simulations. The GA was run up to 3 times on each case with a population size=50, a crossover rate=70\%, a mutation rate=0.95/n$_{\rm bits}$ and a patience before stopping if no progress = 1.5$\times$n$_{\rm bits}$, where n$_{\rm bits}=10\times 2n_{\rm corners}$ for this application. Values in parentheses are those required to reach final solutions.
 The NN was run up to 3 times on each case with the amount of training iteration set to 15, with 64 structures per iteration being added to the training dataset. The four metrics are: the Left-Circularly Polarized reflectivity ($\text{LCP}$), the Right-Circularly Polarized reflectivity ($\text{RCP}$), the Normalized Circular Dichroism $2*|\text{LCP}-\text{RCP}|/(\text{LCP}+\text{RCP})$, and the enhanced fitness function $|\text{LCP}-\text{RCP}|*\max(\text{LCP},\text{RCP})$. 
  The Normalized Circular Dichroism can be interpreted as the relative degree of chirality in the reflection, yielding easily interpretable numbers.}
 \label{Table_comparison}
\end{table}

\subsection{Performance comparison NN and GA optimization pipelines}

The comparison between the NN and GA optimization pipelines highlights their complementary strengths. The GA demonstrates greater robustness when exploring highly complex geometries, whereas the NN approach provides significant computational acceleration, making it particularly suitable for rapid feasibility studies of novel metasurface concepts.

The NN-based pipeline offers high flexibility and adaptability for problems where prior knowledge of the optimal solution is limited. Its architecture can be readily modified to address different levels of complexity, enabling straightforward extension to other optimization tasks. Conversely, the GA is more effective for locating precise optima within a given parameter space once the problem is well defined. As demonstrated in the previous section, the GA outperformed the NN optimization in nearly all cases, albeit at a higher computational cost.

The hybrid optimization strategy proposed here combines the strengths of both methods. Using the NN pipeline for initial, cost-effective screening enables the rapid identification of promising design candidates, which can subsequently be refined through GA optimization. This sequential approach accelerates the overall design process while maintaining accuracy. Moreover, results obtained from both pipelines can serve as an early-stage database for further deep-learning investigations or for transfer learning across related optimization tasks. The combined NN–GA framework can be extended to other inverse-design problems in photonics, including holographic metasurfaces and nonlinear optical devices, offering a scalable and efficient route toward advanced photonic structure design.

Our work builds upon and significantly extends the foundation laid by previous studies that have explored the use of artificial neural networks ANNs and GA, for the optimization of chiral nanostructures. For example, while prior literature has successfully demonstrated the capability of both ANNs and EAs to achieve target chiroptical responses for dielectric metasurfaces \cite{bae2025inverse}, this paper provides a more rigorous and systematic comparative framework, detailing the practical strengths and limitations of both a refined deep learning and a stochastic EA approach for this specific application. Crucially, while other state-of-the-art methods
%\textcolor{red}{[Alex: references are needed here]}
have focused on complex three-dimensional metamaterials \cite{liu2023deep,lee2024advancing,cerniauskas2024machine,garg2025design}, often requiring extensive initial datasets or complex network architectures, our primary contribution lies in the development of a refined optimization pipeline. 

Consequently, our work not only offers a comprehensive comparison of optimization methodologies but also provides an efficient, robust, and manufacturable design recipe for high-performance chiral metasurfaces, validating the power of a systematically optimized machine learning framework in this domain.

\section{Conclusion}

We demonstrated a significantly improved ML-based optimization framework for chiral photonic metasurfaces. 
By combining a refined neural network pipeline with evolutionary strategies, we achieved both high-performance structures and efficient computational scaling. These results promise practical fabrication of chiral mirrors or optical filters and other advanced nanophotonic devices that can be fabricated with different spectral regions and feature scales in mind using top-down or bottom-up lithographic structure definition, ranging from electron-beam to two-photon polymerization writing in photoresist.

\section*{Acknowledgments}
Financial support by the Deutsche Forschungsgemeinschaft (German Research Foundation, Grant No.: DFG RA2841/12-1)---Projektnummer 456700276 (Project ID)---and Chinesisch-Deutsches Zentrum für Wissenschaftsförderung (Sino-German Center for Research Promotion, Grant No.: CDZ GZ1580) is acknowledged. 
A.M. is funded by the Fund for Scientific Research (F.R.S.-FNRS) of Belgium. This research used resources of the “Plateforme Technologique de Calcul Intensif (PTCI)” located at the University of Namur, Belgium, which is supported by the F.R.S.-FNRS under the convention No. 2.5020.11. The PTCI is member of the “Consortium des Equipements de Calcul Intensif (CECI)”. The present research also benefited from computational resources made available on the Tier-1 supercomputer of the Federation Wallonie-Bruxelles, infrastructure funded by the Walloon Region under the grant agreement No. 1117545. 
The authors thank O. Mey for the development of the initial reinforced supervised learning algorithm and technical assistance with the improvement of the existing neural network model. 

\bibliographystyle{unsrt}
\bibliography{ReferencesAdded}

@article{alias2018review,
  title={Review of nanophotonics approaches using nanostructures and nanofabrication for III-nitrides ultraviolet-photonic devices},
  author={Alias, Mohd Sharizal and Tangi, Malleswararao and Holguin-Lerma, Jorge A and Stegenburgs, Edgars and Alatawi, Abdullah A and Ashry, Islam and Subedi, Ram Chandra and Priante, Davide and Shakfa, Mohammad Khaled and Ng, Tien Khee and others},
  journal={Journal of Nanophotonics},
  volume={12},
  number={4},
  pages={043508--043508},
  year={2018},
  publisher={Society of Photo-Optical Instrumentation Engineers}
}

@article{javaid2024reviewing,
  title={Reviewing advances in nanophotonic biosensors},
  author={Javaid, Zunaira and Iqbal, Muhammad Aamir and Javeed, Saher and Maidin, Siti Sarah and Morsy, Kareem and Shati, Ali A and Choi, Jeong Ryeol},
  journal={Frontiers in Chemistry},
  volume={12},
  pages={1449161},
  year={2024},
  publisher={Frontiers Media SA}
}

@article{hughes2018adjoint,
  title={Adjoint method and inverse design for nonlinear nanophotonic devices},
  author={Hughes, Tyler W and Minkov, Momchil and Williamson, Ian AD and Fan, Shanhui},
  journal={ACS Photonics},
  volume={5},
  number={12},
  pages={4781--4787},
  year={2018},
  publisher={ACS Publications}
}

@article{guddala2019valley,
  title={Valley selective optical control of excitons in 2D semiconductors using a chiral metasurface},
  author={Guddala, S and Bushati, R and Li, M and Khanikaev, AB and Menon, VM},
  journal={Optical Materials Express},
  volume={9},
  number={2},
  pages={536--543},
  year={2019},
  publisher={Optical Society of America}
}

@article{im2024perspectives,
  title={Perspectives of chiral nanophotonics: from mechanisms to biomedical applications},
  author={Im, Seongmin and Mousavi, Seyedehniousha and Chen, Yun-Sheng and Zhao, Yang},
  journal={npj Nanophotonics},
  volume={1},
  number={1},
  pages={46},
  year={2024},
  publisher={Nature Publishing Group UK London}
}

@article{kuhner2023unlocking,
  title={Unlocking the out-of-plane dimension for photonic bound states in the continuum to achieve maximum optical chirality},
  author={K{\"u}hner, Lucca and Wendisch, Fedja J and Antonov, Alexander A and B{\"u}rger, Johannes and H{\"u}ttenhofer, Ludwig and de S. Menezes, Leonardo and Maier, Stefan A and Gorkunov, Maxim V and Kivshar, Yuri and Tittl, Andreas},
  journal={Light: Science \& Applications},
  volume={12},
  number={1},
  pages={250},
  year={2023},
  publisher={Nature Publishing Group UK London}
}

@article{li2022empowering,
  title={Empowering metasurfaces with inverse design: principles and applications},
  author={Li, Zhaoyi and Pestourie, Rapha{\"e}l and Lin, Zin and Johnson, Steven G and Capasso, Federico},
  journal={Acs Photonics},
  volume={9},
  number={7},
  pages={2178--2192},
  year={2022},
  publisher={ACS Publications}
}

@article{Wiecha_2021,
  title   = "Deep learning in nano-photonics: inverse design and beyond",
  author  = "Wiecha, P.R. and Arbouet, A. and Girard, C. and Muskens, O.L.",
  journal = "Photonics Res.",
  volume  = "9",
  number  = "5",
  pages   = "B182--B200",
  year    = "2021"
}

@article{hughes2019forward,
  title   = "Forward-mode differentiation of {M}axwell’s equations",
  author  = "Hughes, T.W. and Williamson, I.A.D. and Minkov, M. and Fan, S.",
  journal = "ACS Photonics",
  volume  = "6",
  number  = "11",
  pages   = "3010--3016",
  year    = "2019"
}

@article{minkov2020inverse,
  title   = "Inverse design of photonic crystals through automatic differentiation",
  author  = "Minkov, M. and Williamson, I.A.D. and Andreani, L.C. and Gerace, D. and Lou, B. and Song, A.Y. and Hughes, T.W. and Fan, S.",
  journal = "ACS Photonics",
  volume  = "7",
  number  = "7",
  pages   = "1729--1741",
  year    = "2020"
}

@article{deng2021neural,
  title={Neural-adjoint method for the inverse design of all-dielectric metasurfaces},
  author={Deng, Yang and Ren, Simiao and Fan, Kebin and Malof, Jordan M and Padilla, Willie J},
  journal={Optics Express},
  volume={29},
  number={5},
  pages={7526--7534},
  year={2021},
  publisher={Optical Society of America}
}

@article{ma2019probabilistic,
  title   = "Probabilistic representation and inverse design of metamaterials based on a deep generative model with semi-supervised learning strategy",
  author  = "Ma, Wei and Cheng, Feng and Xu, Yihao and Wen, Qinlong and Liu, Yongmin",
  journal = "Advanced Materials",
  volume  = "31",
  number  = "35",
  pages   = "1901111",
  year    = "2019"
}

@article{Jiang2019,
  title   = "Global optimization of dielectric metasurfaces using a physics driven neural network",
  author  = "Jiang, J. and Fan, J.A.",
  journal = "Nano Letters",
  volume  = "19",
  number  = "",
  pages   = "5366--5372",
  year    = "2019"
}

@article{So_2021,
  title   = "On-demand design of spectrally sensitive multiband absorbers using an artificial neural network",
  author  = "So, S. and Yang, Y. and Lee, T. and Rho, J.",
  journal = "Photonics Res.",
  volume  = "9",
  number  = "4",
  pages   = "B153--B158",
  year    = "2021"
}

@article{So_2023,
  title   = "Revisiting the design strategies for metasurfaces: fundamental physics, optimization, and beyond",
  author  = "So, S. and Mun, J. and Park, J. and Rho, J.",
  journal = "Adv. Mater.",
  volume  = "35",
  number  = "43",
  pages   = "2206399",
  year    = "2023"
}

@article{ma2021deep,
  title={Deep learning for the design of photonic structures},
  author={Ma, Wei and Liu, Zhaocheng and Kudyshev, Zhaxylyk A and Boltasseva, Alexandra and Cai, Wenshan and Liu, Yongmin},
  journal={Nature photonics},
  volume={15},
  number={2},
  pages={77--90},
  year={2021},
  publisher={Nature Publishing Group UK London}
}

@article{Fang2024inverse,
  title   = "Inverse design of lateral hybrid metasurfaces structural colour: an AI approach",
  author  = "Fang, R. and Ghasemi, A. and Zeze, D. and Hedayati, M.K.",
  journal = "RSC Advances",
  volume  = "14",
  number  = "35",
  pages   = "25678--25684",
  year    = "2024"
}

@article{Roy2024pra,
  title={Photonic structures optimization using highly data-efficient deep learning: application to nanofins and annular groove phase masks},
  author={Roy, N. and K\"onig, L. and Absil, O. and Beauthier, Ch. and Mayer, A. and Lobet, M.},
  journal={Phys. Rev. A},
  volume={109},
  number={},
  pages={013514},
  year={2024},
  publisher={American Physical Society}
}

@article{Christensen_2020,
  title   = "Predictive and generative machine learning models for photonic crystals",
  author  = "Christensen, T. and Loh, C. and Picek, S. and Jakobovic, D. and Jing, L. and Fisher, S. and Ceperic, V. and Joannopoulos, J.D. and Soljacic, M.",
  journal = "Nanophotonics",
  volume  = "9",
  number  = "13",
  pages   = "4183--4192",
  year    = "2020"
}

@article{Chen2020PINN,
  title   = "Physics-informed neural networks for inverse problems in nano-optics and metamaterials",
  author  = "Chen, Y. and Lu L. and Karniadakis, G.E. and Negro, L.D.",
  journal = "Optics Express",
  volume  = "28",
  number  = "8",
  pages   = "11618--11633",
  year    = "2020"
}

@article{Lim2022MaxwellNet,
  title   = "MaxwellNet: Physics-driven deep neural network training based on Maxwell’s equations",
  author  = "Lim, J. and Psaltis, D.",
  journal = "APL Photonics",
  volume  = "7",
  number  = "1",
  pages   = "011301",
  year    = "2022"
}

@article{Deng2025,
  title   = "Physics-informed learning in artificial electromagnetic materials",
  author  = "Deng, Y. and Fan, K. and Jin, B. and Malof, J. and Padilla, W.J.",
  journal = "Appl. Phys. Rev.",
  volume  = "12",
  number  = "1",
  pages   = "011331",
  year    = "2025"
}

@article{LuJordan2025agentic,
  title={An Agentic Framework for Autonomous Metamaterial Modeling and Inverse Design},
  author={LuJordan, D. and Malof, M. and Padilla, W.J.},
  journal={ACS Photonics},
  volume={12},
  number={11},
  pages={6071--6080},
  year={2025},
  publisher={American Chemical Society}
}

@article{mey2022machine,
  title={Machine learning-based optimization of chiral photonic nanostructures: evolution-and neural network-based designs},
  author={Mey, Oliver and Rahimi-Iman, Arash},
  journal={physica status solidi (RRL)--Rapid Research Letters},
  volume={16},
  number={2},
  pages={2100571},
  year={2022},
  publisher={Wiley Online Library}
}

@unpublished{Khepri,
    author = {Nicolas Roy} ,
    title = {Khepri: Kode for High-Efficiency Propagation of Radiant Interactions},
    note = {https://github.com/Kaeryv/Khepri}
}

@article{Moharam_1981,
  author  = "Moharam, M. and Gaylord, T.",
  title   = "Rigorous coupled-wave analysis of planar-grating diffraction",
  journal = "J. Opt. Soc. Am. A",
  volume  = "71",
  number  = "7",
  pages   = "811--818",
  year    = "1981"
}

@article{RCWA,
  title={New formulation of the Fourier modal method for crossed surface-relief gratings},
  author={Li, Lifeng},
  journal={Journal of the Optical Society of America A},
  volume={14},
  number={10},
  pages={2758--2767},
  year={1997},
  publisher={Optical Society of America}
}

@article{S4,
  title={S4: A free electromagnetic solver for layered periodic structures},
  author={Liu, Victor and Fan, Shanhui},
  journal={Computer Physics Communications},
  volume={183},
  number={10},
  pages={2233--2244},
  year={2012},
  publisher={Elsevier}
}

@article{aspnes1983dielectric,
  title={Dielectric functions and optical parameters of si, ge, gap, gaas, gasb, inp, inas, and insb from 1.5 to 6.0 ev},
  author={Aspnes, David E and Studna, AA},
  journal={Physical review B},
  volume={27},
  number={2},
  pages={985},
  year={1983},
  publisher={APS}
}

@article{Mayer_SPIE2018,
  author  = "Mayer, A. and Lobet, M.",
  title   = "{UV} to near-infrared broadband pyramidal absorbers via a genetic algorithm optimization approach",
  journal = "Proc. SPIE",
  volume  = "10671",
  pages   = "1067127",
  year    = "2018"
}

@article{Mayer_SPIE2020,
  author  = "Mayer, A. and Griesse-Nascimento, S. and Bi, H. and Mazur, E. and Lobet, M.",
  title   = "Optimization by a genetic algorithm of pyramidal structures made of one, two or three stacks of metal/dielectric layers for a quasi-perfect broadband absorption of UV to near-infrared radiations",
  journal = "Proc. SPIE",
  volume  = "11344",
  pages   = "113441L",
  year    = "2020"
}

@article{Mayer_OE2022,
  author  = "Mayer, A. and Bi, H. and Griesse-Nascimento, S. and Hackens, B. and Loicq, J. and Mazur, E. and Deparis, O. and Lobet, M.",
  title   = "Genetic-algorithm-aided ultra-broadband perfect absorbers using plasmonic metamaterials",
  journal = "Opt. Express",
  volume  = "30",
  number  = "2",
  pages   = "1167--1181",
  year    = "2022"
}

@article{Mayer_SPIE2024,
  title={Optimization by a genetic algorithm of nanopyramidal broadband quasi-perfect absorbers with deeper insight into the stability of optimal solutions},
  author={Mayer, A. and Deparis, O. and Lobet, M.},
  journal={Proc. SPIE},
  volume={13017},
  pages={1301714},
  year={2024},
  publisher={SPIE}
}

@book{Eiben_2007,
  author    = "Eiben, A.E. and Smith, J.E.",
  title     = "Introduction to Evolutionary Computing",
  publisher = "Springer-Verlag",
  address   = "Berlin",
  edition   = "second",
  year      = "2007"
}

@article{bai2007optical,
  title={Optical activity in planar chiral metamaterials: Theoretical study},
  author={Bai, Benfeng and Svirko, Yuri and Turunen, Jari and Vallius, Tuomas},
  journal={Physical Review A—Atomic, Molecular, and Optical Physics},
  volume={76},
  number={2},
  pages={023811},
  year={2007},
  publisher={APS}
}

@article{bae2025inverse,
  title={Inverse design for enhanced chiroptical response with chiral nanophotonic structures},
  author={Bae, Munseong and Pan, Chia-Chun and Kang, Chanik and Bae, Jinseong and Park, Donghyun and Lee, Seokho and Park, Cherry and Ren, Haoran and Rho, Junsuk and Chung, Haejun and others},
  journal={APL Photonics},
  volume={10},
  number={10},
  year={2025},
  publisher={AIP Publishing}
}

@article{liu2023deep,
  title={Deep learning for the design of phononic crystals and elastic metamaterials},
  author={Liu, Chen-Xu and Yu, Gui-Lan},
  journal={Journal of Computational Design and Engineering},
  volume={10},
  number={2},
  pages={602--614},
  year={2023},
  publisher={Oxford University Press}
}

@article{lee2024advancing,
  title={Advancing programmable metamaterials through machine learning-driven buckling strength optimization},
  author={Lee, Sangryun and Kwon, Junpyo and Kim, Hyunjun and Ritchie, Robert O and Gu, Grace X},
  journal={Current Opinion in Solid State and Materials Science},
  volume={31},
  pages={101161},
  year={2024},
  publisher={Elsevier}
}

@article{cerniauskas2024machine,
  title={Machine intelligence in metamaterials design: a review},
  author={Cerniauskas, Gabrielis and Sadia, Haleema and Alam, Parvez},
  journal={Oxford Open Materials Science},
  volume={4},
  number={1},
  pages={itae001},
  year={2024},
  publisher={Oxford University Press}
}

@article{garg2025design,
  title={Design and Optimization of a Metamaterial Absorber Using Machine Learning Models: P. Garg},
  author={Garg, Priyanka},
  journal={Journal of Electronic Materials},
  pages={1--10},
  year={2025},
  publisher={Springer}
}

\appendix
\section*{Appendix}

\begin{figure}[ht]
 \centering
 \includegraphics[width=0.49\linewidth]{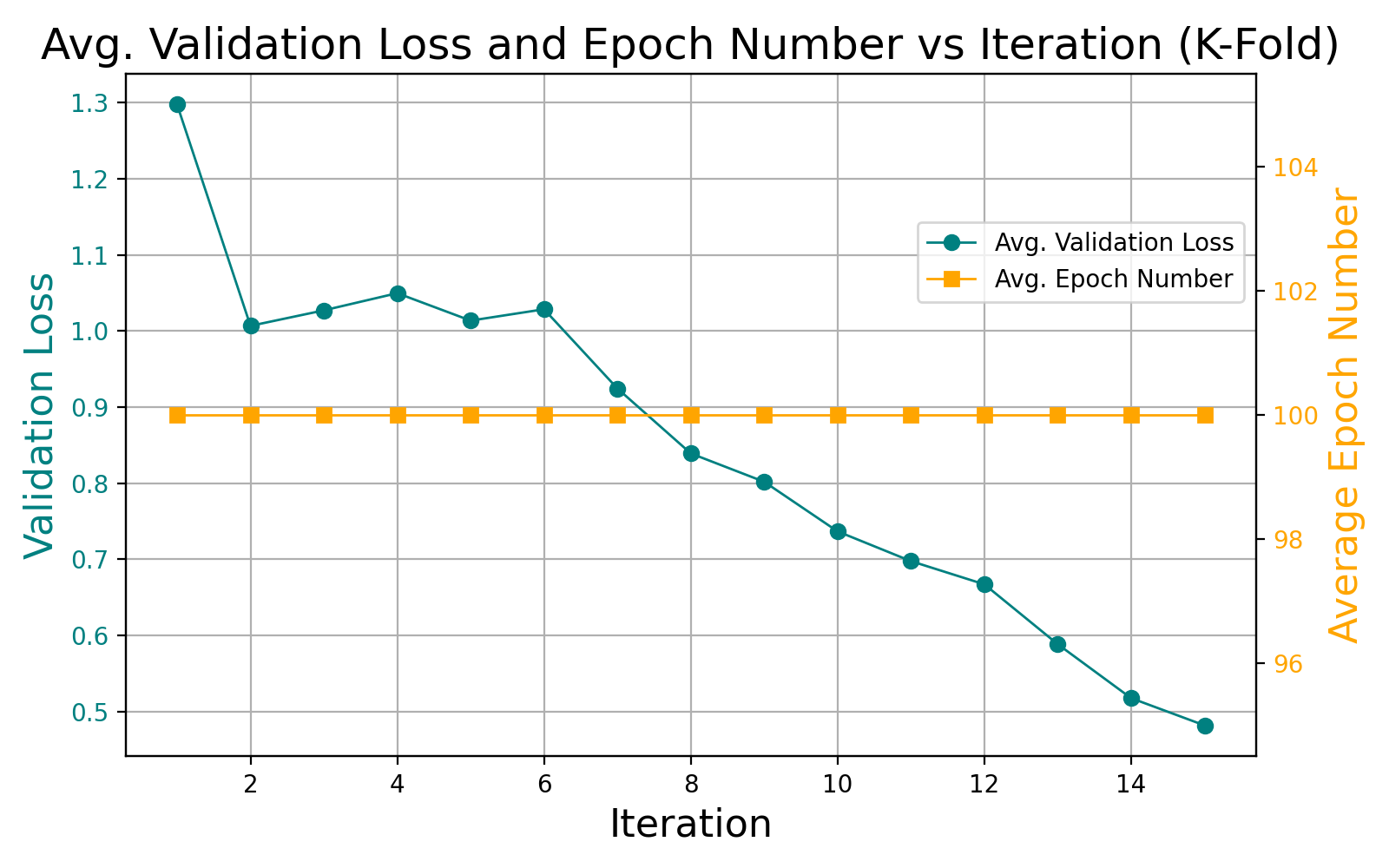}
 \includegraphics[width=0.49\linewidth]{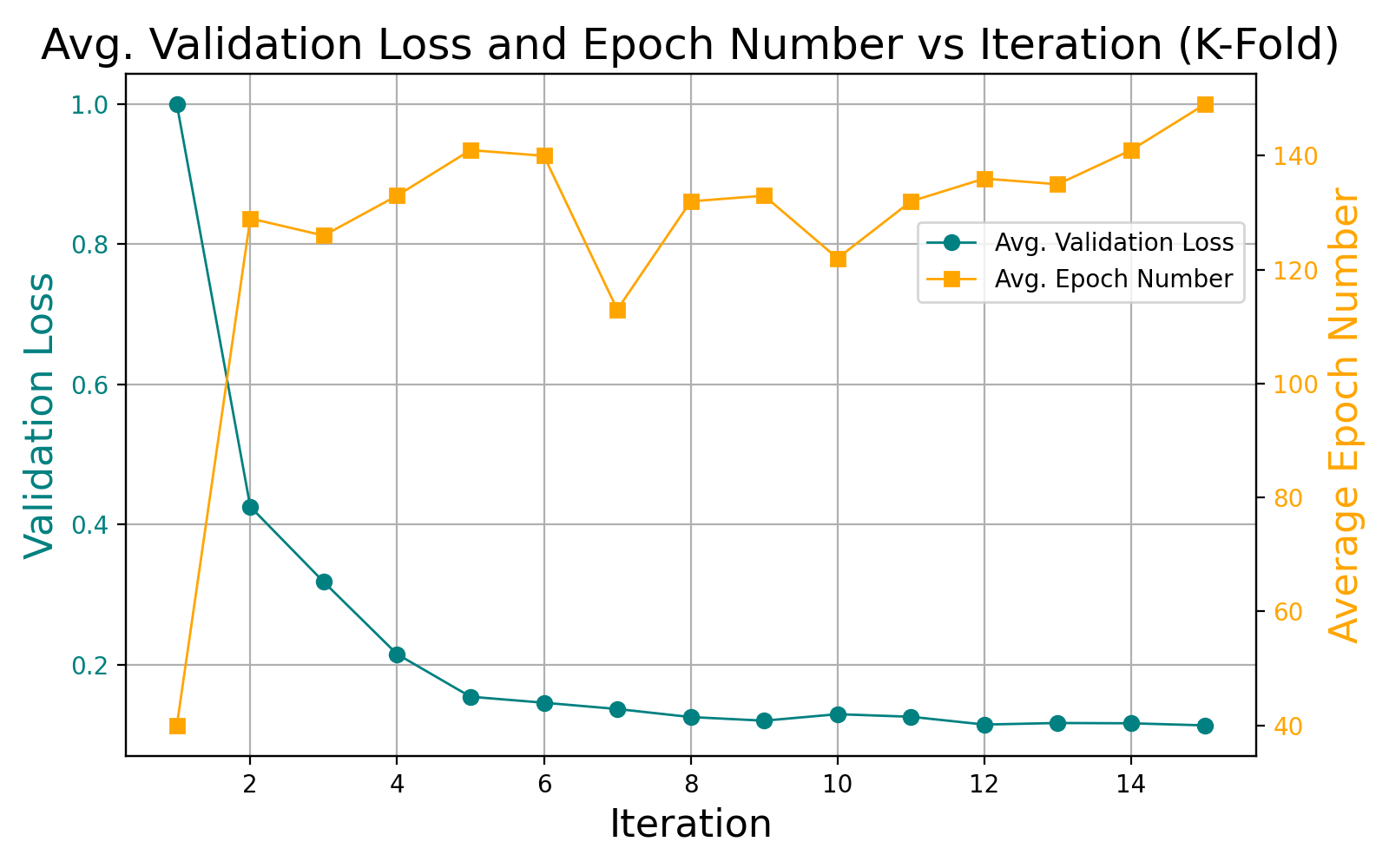}
 \caption{Left: validation loss for the 15 iterations of the optimization pipeline using the 1-output-neuron neural network (left)
 or the 2-output-neurons neural network with improved epoch selection methods, fitness function and geometry augmentation (right).}
 \label{fig:Validation}
\end{figure}

Figure \ref{fig:Validation} illustrates the convergence behavior of the neural network (NN) optimization pipeline by comparing the Average Validation Loss and the Average Epoch Number per iteration for two different network architectures.
The left panel shows the optimization results using the initial 1-output-neuron NN, $\Delta R_{\rm CD}$. In this case, the average validation loss remains high, ending near 0.5. This high loss is attributed to underfitting, as the simple network could not effectively learn the complex nonlinear relationships, and the average epoch number was fixed. The right panel shows the optimization using the advanced 2-output-neurons NN, which simultaneously predicts $\Delta R_{\rm CD}$ and $R_{\rm pref}$. The validation loss for this improved network architecture drops sharply, stabilizing at approximately 0.1 by the final iteration—a five-fold reduction, indicating superior model accuracy and generalization. This improvement is partly due to the adaptive epoch selection strategy, which adjusts the training duration to prevent both overfitting and underfitting, reflected by the varying average epoch number.

\begin{figure}[H]
 \centering
 \includegraphics[width=1\linewidth]{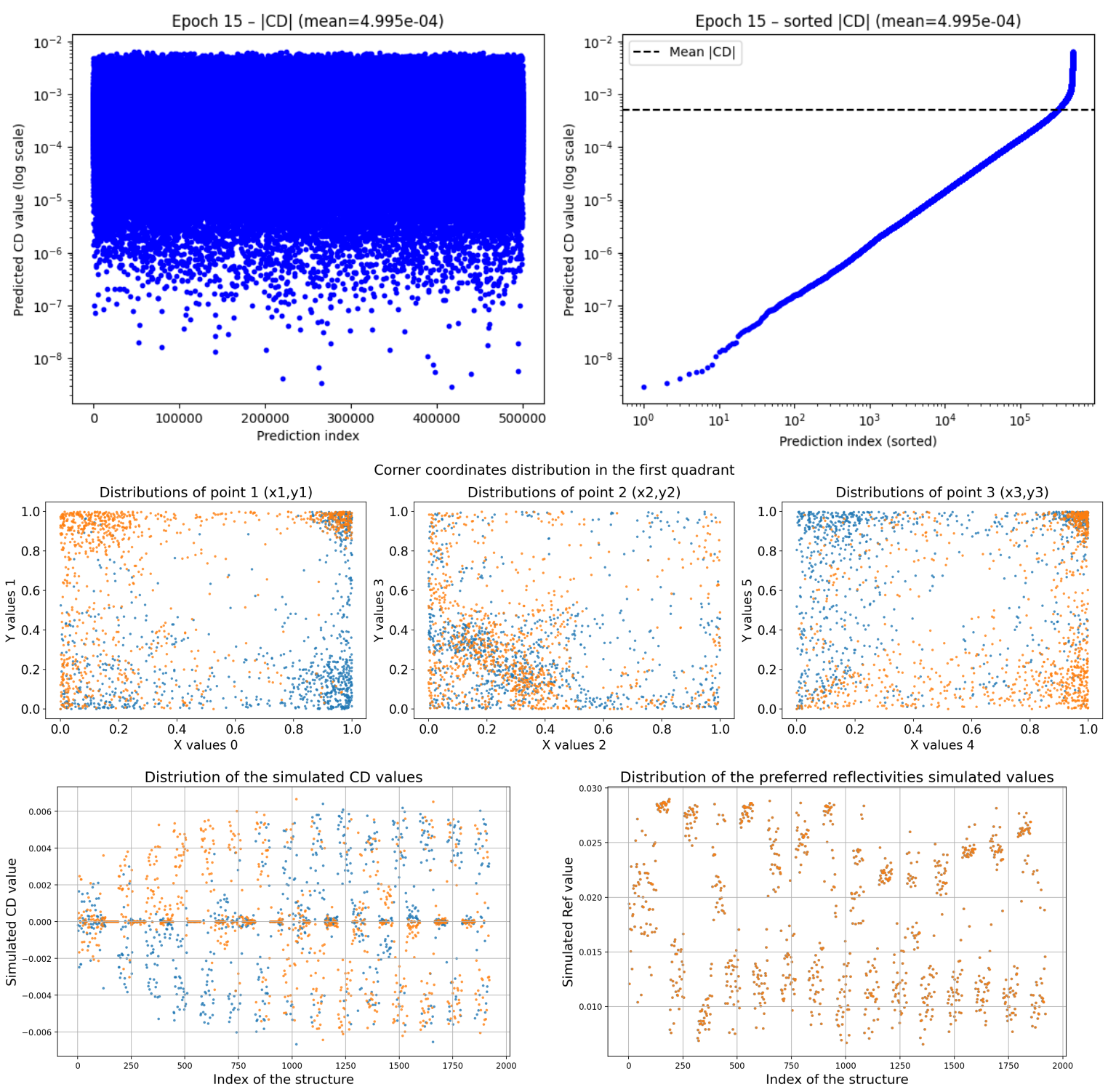}
 \caption{On the top the inference results on the last epoch are presented, displaying how only a small part of the structures show a high CD value. On the bottom, the dataset obtained at the end of the optimization is presented, showing the effect of the symmetry augmentation (in blue: original structures; in orange: structures obtained by augmentation).}
 \label{fig:Dataset}
\end{figure}

Figure \ref{fig:Dataset} provides a visual breakdown of the data exploration and augmentation strategy. The top panels show the inference results on the last epoch, demonstrating that even among 500,000 structures per iteration, only a small portion exhibits a high predicted  $\Delta R_{\rm CD}$ value. The bottom panels illustrate the composition of the final training dataset and the effect of the geometric data augmentation. This strategy exploits the mirror symmetry between enantiomeric chiral metasurfaces. By simply swapping the LCP and RCP labels for each simulated sample (an approach consistent with the underlying physics), the training set is effectively doubled without requiring any additional, computationally expensive simulations. The original simulated structures are shown in blue, and the augmented (enantiomeric) structures are shown in orange, ensuring a more balanced representation across the design space for both $\Delta R_{\rm CD}$ and $R_{\rm pref}$.

\end{document}